\documentstyle[12pt]{article}

\advance \topmargin by -\headheight
\advance \topmargin by -\headsep
     
\evensidemargin \oddsidemargin
\begin{document}
\newcommand{\sect}[1]{\setcounter{equation}{0}\section{#1}}
\renewcommand{\theequation}{\thesection.\arabic{equation}}
\date{} 
\topmargin -.6in
\def\nonu{\nonumber}
\def\rf#1{(\ref{eq:#1})}
\def\lab#1{\label{eq:#1}} 
\def\br{\begin{eqnarray}}
\def\er{\end{eqnarray}}
\def\be{\begin{equation}}
\def\ee{\end{equation}}
\def\0{\nonumber}
\def\lb{\lbrack}
\def\rb{\rbrack}
\def\({\left(}
\def\){\right)}
\def\v{\vert}
\def\bv{\bigm\vert}
\def\lskip{\vskip\baselineskip\vskip-\parskip\noindent}
\relax
\newcommand{\nit}{\noindent}
\newcommand{\ct}[1]{\cite{#1}}
\newcommand{\bi}[1]{\bibitem{#1}}
\def\a{\alpha}
\def\b{\beta}
\def\ca{{\cal A}}
\def\cm{{\cal M}}
\def\cn{{\cal N}}
\def\cf{{\cal F}}
\def\d{\delta} 
\def\D{\Delta}
\def\eps{\epsilon}
\def\g{\gamma}
\def\G{\Gamma}
\def\grad{\nabla}
\def\h{ {1\over 2}  }
\def\hc{\hat{c}}
\def\hd{\hat{d}}
\def\hg{\hat{g}}
\def\hp{ {+{1\over 2}}  }
\def\hm{ {-{1\over 2}}  }
\def\k{\kappa}
\def\l{\lambda}
\def\L{\Lambda}
\def\lg{\langle}
\def\m{\mu}
\def\n{\nu}
\def\o{\over}
\def\om{\omega}
\def\O{\Omega}
\def\p{\phi}
\def\pa{\partial}
\def\pr{\prime}
\def\ra{\rightarrow}
\def\rh{\rho}
\def\rg{\rangle}
\def\s{\sigma}
\def\t{\tau}
\def\th{\theta}
\def\ti{\tilde}
\def\wti{\widetilde}
\def\inte{\int dx }
\def\xb{\bar{x}}
\def\yb{\bar{y}}

\def\tr{\mathop{\rm tr}}
\def\Tr{\mathop{\rm Tr}}
\def\partder#1#2{{\partial #1\over\partial #2}}
\def\ds{{\cal D}_s}
\def\wtwo{{\wti W}_2}
\def\lie{{\cal G}}
\def\alie{{\widehat \lie}}
\def\dlie{{\cal G}^{\ast}}
\def\elie{{\widetilde \lie}}
\def\edlie{{\elie}^{\ast}}
\def\hlie{{\cal H}}
\def\wlie{{\widetilde \lie}}

\def\rlx{\relax\leavevmode}
\def\inbar{\vrule height1.5ex width.4pt depth0pt}
\def\IZ{\rlx\hbox{\sf Z\kern-.4em Z}}
\def\IR{\rlx\hbox{\rm I\kern-.18em R}}
\def\IC{\rlx\hbox{\,$\inbar\kern-.3em{\rm C}$}}
\def\one{\hbox{{1}\kern-.25em\hbox{l}}}

\def\PRL#1#2#3{{\sl Phys. Rev. Lett.} {\bf#1} (#2) #3}
\def\NPB#1#2#3{{\sl Nucl. Phys.} {\bf B#1} (#2) #3}
\def\NPBFS#1#2#3#4{{\sl Nucl. Phys.} {\bf B#2} [FS#1] (#3) #4}
\def\CMP#1#2#3{{\sl Commun. Math. Phys.} {\bf #1} (#2) #3}
\def\PRD#1#2#3{{\sl Phys. Rev.} {\bf D#1} (#2) #3}
\def\PLA#1#2#3{{\sl Phys. Lett.} {\bf #1A} (#2) #3}
\def\PLB#1#2#3{{\sl Phys. Lett.} {\bf #1B} (#2) #3}
\def\JMP#1#2#3{{\sl J. Math. Phys.} {\bf #1} (#2) #3}
\def\PTP#1#2#3{{\sl Prog. Theor. Phys.} {\bf #1} (#2) #3}
\def\SPTP#1#2#3{{\sl Suppl. Prog. Theor. Phys.} {\bf #1} (#2) #3}
\def\AoP#1#2#3{{\sl Ann. of Phys.} {\bf #1} (#2) #3}
\def\PNAS#1#2#3{{\sl Proc. Natl. Acad. Sci. USA} {\bf #1} (#2) #3}
\def\RMP#1#2#3{{\sl Rev. Mod. Phys.} {\bf #1} (#2) #3}
\def\PR#1#2#3{{\sl Phys. Reports} {\bf #1} (#2) #3}
\def\AoM#1#2#3{{\sl Ann. of Math.} {\bf #1} (#2) #3}
\def\UMN#1#2#3{{\sl Usp. Mat. Nauk} {\bf #1} (#2) #3}
\def\FAP#1#2#3{{\sl Funkt. Anal. Prilozheniya} {\bf #1} (#2) #3}
\def\FAaIA#1#2#3{{\sl Functional Analysis and Its Application} {\bf #1} (#2)
#3}
\def\BAMS#1#2#3{{\sl Bull. Am. Math. Soc.} {\bf #1} (#2) #3}
\def\TAMS#1#2#3{{\sl Trans. Am. Math. Soc.} {\bf #1} (#2) #3}
\def\InvM#1#2#3{{\sl Invent. Math.} {\bf #1} (#2) #3}
\def\LMP#1#2#3{{\sl Letters in Math. Phys.} {\bf #1} (#2) #3}
\def\IJMPA#1#2#3{{\sl Int. J. Mod. Phys.} {\bf A#1} (#2) #3}
\def\AdM#1#2#3{{\sl Advances in Math.} {\bf #1} (#2) #3}
\def\RMaP#1#2#3{{\sl Reports on Math. Phys.} {\bf #1} (#2) #3}
\def\IJM#1#2#3{{\sl Ill. J. Math.} {\bf #1} (#2) #3}
\def\APP#1#2#3{{\sl Acta Phys. Polon.} {\bf #1} (#2) #3}
\def\TMP#1#2#3{{\sl Theor. Mat. Phys.} {\bf #1} (#2) #3}
\def\JPA#1#2#3{{\sl J. Physics} {\bf A#1} (#2) #3}
\def\JSM#1#2#3{{\sl J. Soviet Math.} {\bf #1} (#2) #3}
\def\MPLA#1#2#3{{\sl Mod. Phys. Lett.} {\bf A#1} (#2) #3}
\def\JETP#1#2#3{{\sl Sov. Phys. JETP} {\bf #1} (#2) #3}
\def\JETPL#1#2#3{{\sl  Sov. Phys. JETP Lett.} {\bf #1} (#2) #3}
\def\PHSA#1#2#3{{\sl Physica} {\bf A#1} (#2) #3}

\newcommand\twomat[4]{\left(\begin{array}{cc}  
{#1} & {#2} \\ {#3} & {#4} \end{array} \right)}
\newcommand\twocol[2]{\left(\begin{array}{cc}  
{#1} \\ {#2} \end{array} \right)}
\newcommand\twovec[2]{\left(\begin{array}{cc}  
{#1} & {#2} \end{array} \right)}

\newcommand\threemat[9]{\left(\begin{array}{ccc}  
{#1} & {#2} & {#3}\\ {#4} & {#5} & {#6}\\ {#7} & {#8} & {#9} \end{array} \right)}
\newcommand\threecol[3]{\left(\begin{array}{ccc}  
{#1} \\ {#2} \\ {#3}\end{array} \right)}
\newcommand\threevec[3]{\left(\begin{array}{ccc}  
{#1} & {#2} & {#3}\end{array} \right)}

\newcommand\fourcol[4]{\left(\begin{array}{cccc}  
{#1} \\ {#2} \\ {#3} \\ {#4} \end{array} \right)}
\newcommand\fourvec[4]{\left(\begin{array}{cccc}  
{#1} & {#2} & {#3} & {#4} \end{array} \right)}

\begin{titlepage}
\vspace*{-2 cm}
\noindent


\vskip 1 cm
\begin{center}
{\Large\bf Non Abelian Toda models and the Constrained KP hierarchies}  \footnote{talk given at 7th International
 Wigner Symposium, College Park, Maryland,  August 2001} \vglue 1  true cm
I. Cabrera-Carnero\footnote{cabrera@ift.unesp.br}, { J.F. Gomes}\footnote{jfg@ift.unesp.br}, 
E.P. Gueuvoghlanian\footnote{gueuvogh@ift.unesp.br}
 { G.M. Sotkov}\footnote{sotkov@ift.unesp.br} and { A.H. Zimerman}\footnote{zimerman@ift.unesp.br}\\

\vspace{1 cm}

{\footnotesize Instituto de F\'\i sica Te\'orica - IFT/UNESP\\
Rua Pamplona 145\\
01405-900, S\~ao Paulo - SP, Brazil}\\

\vspace{1 cm}

\end{center}

\normalsize
\vskip 0.2cm

\begin{center}
{\large {\bf ABSTRACT}}\\
\end{center}
\noindent

A general construction of affine  Non Abelian Toda models  in terms of gauged
two loop WZNW model is discussed. Its connection to non relativistic models
 corresponding to constrained KP hierarchies is
established in terms of time evolution associated  to 
 positive and negative  grading of the Lie algebra.

\noindent

\vglue 1 true cm

\end{titlepage}

\sect{Introduction} 

The abelian affine Toda field theories provide a large class of relativistic invariant 
integrable models in two dimensions
associated to an affine Lie algebra $\tilde \lie$ (loop algebra) admiting  solitons solutions.  The
Toda fields are defined to parametrize a finite dimensional  abelian manifold (Cartan subalgebra of  
$\tilde \lie$) and their solitonic character (in the imaginary coupling regime \cite{holl}) 
is a consequence of the infinite dimensional Lie
algebraic structure responsible for the multivacua configuration leading to a nontrivial topological
structure.

A more general class of affine Toda models is obtained by introducing a non abelian structure to the
abelian manifold (Cartan subalgebra of  
$\tilde \lie$) parametrized by the Toda fields. 
 A systematic  manner in classifying the Toda models
\cite{lez-sav}
is in terms of a grading operator $Q$ that decomposes the affine Lie algebra $\tilde \lie = \oplus
\lie_i$, where the graded subspaces are defined by $[Q, \lie_i ] = i \lie_i$.  The Toda fields are
defined to parametrize the zero grade subspace $\lie_0 \subset \lie$.  

 The simplest model 
is obtained when  $\lie = \tilde {SL}(2),$ is
decomposed according to the homogeneous gradation leading to the Lund-Regge model. 
 For $SL(r +1), $ a general 
construction  is discussed in ref. \cite{dyonic} leading to actions
 corresponding to Lund-Regge interacting with abelian Toda
models.  
In this note we discuss a systematic construction  of    non abelian Toda models
associated to $\lie_0 = SL(2)^p \otimes U(1)^{rank \lie - p}$.  
The simplest non trivial generalization of the Lund-Regge model is obtained
 $\lie_0 = SL(2) \otimes SL(2) \otimes U(1) \subset \tilde {SL}(4)$. 
  The  action, equations of motion and zero curvature
 representation are explicitly constructed.  
 The class of relativistic invariant non abelian Toda models share a common
 algebraic structure with the constrained KP type models \cite{jmp}. 
  These are non relativistic systems whose multi time 
  equation of motion generate an integrable hierarchy  associated to positive grading. 
It was shown in \cite{jpa7} that the Lund-Regge  and the non linear Schroedinger models 
correspond to the same hierarchy when time evolution is also assign to negative grade generators.  
The problem of  constructing integrable hierarchies associated to positive 
and negative gradings was further investigated and generalized to $SL(r+1)$ in
\cite{rh} where it was shown to be connected to the Riemann-Hilbert problem.
Here, several  examples for $SL(2), SL(3) $ and $SL(4)$ are discussed in detail.  In particular 
 the soliton solutions within positive and negative hierarchies are shown to be related.

\sect{Construction of the Model}

The generic NA Toda models  are classified
 according to a $\lie_0 \subset \lie$ embedding  induced
by the grading operator $Q$ decomposing an finite or infinite dimensional Lie algebra 
$\lie = \oplus _{i} \lie _i $ where $
[Q,\lie_{i}]=i\lie_{i}$ and $ [\lie_{i},\lie_{j}]\subset \lie_{i+j}$.  A group
element $g$ can then be written in terms of the Gauss decomposition as 
\be
g= NBM
\label{1}
\ee
where $N=\exp \lie_< $, $B=\exp \lie_{0} $ and
$M=\exp \lie_> $.  The physical fields lie in the zero grade subgroup $B$ 
and the
models we seek correspond to the coset $H_- \backslash G/H_+ $, for $H_{\pm} $ generated by
positive/negative  grade operators.

For consistency with the hamiltonian reduction formalism, the phase space of
the G-invariant WZNW model is  reduced by specifying the constant
generators $\eps_{\pm}$ of grade $\pm 1$.  In order to derive 
 an action for $B  $,  invariant under 
\begin{eqnarray}
g\longrightarrow g^{\prime}=\alpha_{-}g\alpha_{+},
\label{2}
\end{eqnarray}
where $\a_{\pm}(z, \bar z)$ lie in the positive/negative grade subgroup
 we have to introduce a set of  auxiliary
gauge fields $A \in \lie _{<} $ and $\bar A \in \lie _{>}$ transforming as 
\begin{eqnarray}
A\longrightarrow A^{\prime}=\alpha_{-}A\alpha_{-}^{-1}
+\alpha_{-}\partial \alpha_{-}^{-1},
\quad \quad 
\bar{A}\longrightarrow \bar{A}^{\prime}=\alpha_{+}^{-1}\bar{A}\alpha_{+}
+\bar{\partial}\alpha_{+}^{-1}\alpha_{+},
\label{3}
\end{eqnarray}
where $z = {1\o 2} (t+x), \bar z = {1\o 2} (t-x)$.
 The resulting action is the $G/H (= H_- \backslash G/H_+ )$
 gauged WZNW    
\begin{eqnarray}
S_{G/H}(g,A,\bar{A})&=&S_{WZNW}(g)
\nonumber
\\
&-&\frac{k}{4\pi}\int d^2x Tr\( A(\bar{\partial}gg^{-1}-\epsilon_{+})
+\bar{A}(g^{-1}\partial g-\epsilon_{-})+Ag\bar{A}g^{-1}\) .
\nonumber
\end{eqnarray}
Since the action $S_{G/H}$ is $H$-invariant,
 we may choose $\alpha_{-}=N_{}^{-1}$
and $\alpha_{+}=M_{}^{-1}$. From the orthogonality  of the graded 
subpaces, i.e. $Tr( \lie _i\lie _j ) =0, i+j \neq 0$, we find
\begin{eqnarray}
S_{G/H}(g,A,\bar{A})&=&S_{G/H}(B,A^{\prime},\bar{A}^{\prime})
\nonumber
\\
&=&S_{WZNW}(B)-\frac{k}{4\pi}
\int d^2x Tr[-A^{\prime}\epsilon_{+}-\bar{A}^{\prime}\epsilon_{-}
+A^{\prime}B\bar{A}^{\prime}B^{-1}],
\label{14}
\end{eqnarray}
where 
\begin{eqnarray}
S_{WZNW}=- \frac{k}{4\pi }\int d^2xTr(g^{-1}\partial gg^{-1}\bar{\partial }g)
+\frac{k}{24\pi }\int_{D}\epsilon^{ijk}
Tr(g^{-1}\partial_{i}gg^{-1}\partial_{j}gg^{-1}\partial_{k}g)d^3x,
\label{3a}
\end{eqnarray}
and the topological term denotes a surface integral  over a ball $D$
identified as  space-time.

Action (\ref{14}) describes the non singular Toda models among which we find the
Conformal and the Affine abelian Toda models where 
$Q=\sum_{i=1}^{r}\frac{2\lambda_{i}\cdot H}{\alpha_{i}^{2}}, \quad 
 \epsilon_{\pm}=\sum_{i=1}^{r} \mu_{\pm i}E_{\pm \alpha_{i}}$ and 
$Q= h d + \sum_{i=1}^{r}\frac{2\lambda_{i}\cdot H^{(0)}}{\alpha_{i}^{2}}, \quad 
\epsilon_{\pm}=\sum_{i=1}^{r} \mu_{\pm i}E_{\pm \alpha_{i}}^{(0)} + 
\mu_{0}E_{\mp \psi }^{(\pm 1)}$
respectively,  $\psi $ denotes the highest root,  $\lambda_i$ are the 
fundamental weights  and $h =  1+ \sum_{i=1}^{r}\psi \cdot \l_i $ is 
the coxeter number of $\lie $. 

Performing the integration  over the auxiliary fields $A$ and $\bar A$, 
the functional integral 
\be
 Z_{\pm}=\int DAD\bar{A}\exp (F_{\pm}),
\label{fpm}
\ee 
 where 
 \be
F_{\pm} = {-{k\o {2\pi}}}\int \(Tr
 (A - B {\eps_-} B^{-1})B(\bar A - B^{-1} {\eps_+} B) B^{-1}\)d^2x
\label{fmm}
\ee
yields the effective action
\be
S = S_{WZNW} (B) + {{k\o {2\pi}}} \int Tr \( \eps_+ B  \eps_- B^{-1}\)d^2x
\label{spm}
\ee
The action (\ref{spm}) describes integrable perturbations of the $\lie_0$-WZNW model. 
 Those perturbations are classified in
terms of the possible constant grade $\pm 1$ operators $\eps _{\pm}$.

 More interesting cases 
arises in
connection with non abelian embeddings $\lie_0 \subset \lie $.  In particular, if we 
supress $p$
alternate fundamental weights from $Q$, the zero grade subspace $\lie_0$,
acquires a nonabelian 
structure
$sl(2)^p \otimes u(1)^{rank \lie -p}$.  Let us consider for instance 
$Q= h^{\pr} d + \sum_{i\neq 1,3, \cdots }^{r}\frac{2\lambda_{i}\cdot H}{\alpha_{i}^{2}}$, 
where $h^{\pr} =0$ or
$h^{\pr} \neq 0$ \footnote{ For the Kac-Moody case we are suppressing the
index $(0)$ in defining the Cartan subalgebra of $\lie $} corresponding to  the
Conformal or Affine  nonabelian (NA) Toda  respectively. The absence of
$\lambda_{al}, al= 1, 3, \cdots $ in $Q$ 
 prevents the contribution of the simple root step operator
$E_{\a_{al}}^{(0)}$ in constructing the grade one operator $\eps_+$. 
It in fact, allows for  reducing
the phase space  even further.  This fact can be understood by enforcing the 
nonlocal constraint $J_{Y^l \cdot H} = \bar J_{Y^l \cdot H} = 0$
where $Y^l, l=1, \cdots t \leq p $ is such that $[Y^l\cdot H , \eps_{\pm}] = 0$ and 
$J=g^{-1}\partial g$ and $\bar{J}=-\bar{\partial}gg^{-1}$.  Those generators 
of $\lie_0$ commuting with
$\eps_{\pm}$ define a subalgebra $\lie_0^0 \subset \lie_0 $.
 Such subsidiary constraint is incorporated into the action by
  requiring symmetry under  
\begin{eqnarray}
g\longrightarrow g^{\prime}=\alpha_{0}g\alpha_{0}'
\label{5}
\end{eqnarray} 
where we shall consider   $\a^{\pr}_{0}
 =\alpha_{0}(z, \bar z) \in \lie_0^0 $, i.e., {\it axial symmetry} (the 
 {\it vector }
 gauging is obtained by choosing $\a^{\pr}_{0}
 ={\alpha_{0}}^{-1}(z, \bar z) \in \lie_0^0 $, see for instance  \cite{wigner99}). 
Auxiliary gauge fields $A_0 = \sum_{l=1}^{t}a_0^l Y^l\cdot H$ and $\bar A_0= 
\sum_{l=1}^{t}\bar a_0^l Y^l\cdot H\in  \lie_{0}^{0}$ 
transforming as 
\begin{eqnarray}
A_{0}\longrightarrow A_{0}^{\prime}=A_{0}-\alpha_{0}^{-1}\partial \alpha_{0},
\quad \quad 
\bar{A}_{0}\longrightarrow \bar{A}_{0}^{\prime}=\bar{A}_{0}
- \bar{\partial}\alpha_{0}^{\pr}(\alpha_{0}^{\pr})^{-1}.
\nonu 
\end{eqnarray}
 are 
introduced to construct  an invariant action under transformations  (\ref{5}) 
\br 
S(B,{A}_{0},\bar{A}_{0} ) &=& S(g_0^f,{A^{\pr}}_{0},\bar{A^{\pr}}_{0} )  
 = S_{WZNW}(B)+ 
 {{k\o {2\pi}}} \int Tr \( \eps_+ B  \eps_- B^{-1}\) d^2x\nonu \\   
  &-&{{k\o {2\pi}}}\int Tr\(  A_{0}\bar{\partial}B
B^{-1} + \bar{A}_{0}B^{-1}\partial B
+ A_{0}B\bar{A}_{0}B^{-1} + A_{0}\bar{A}_{0} \)d^2x \nonu \\
\label{aa}
\er
Such residual gauge symmetry allows us to eliminate   extra fields
 associated to $Y^l\cdot H, l=1, \cdots ,t \leq p$.
  Notice that the physical fields
 $g_0^f$ lie in the coset $\lie_0 /{\lie_0^0} = ({{sl(2)^{p}\otimes
u(1)^{rank \lie -p}})/u(1)^t}$  and are classified according to the 
gradation $Q$.  It therefore follows that $S(B,A_0,\bar{A_0})=
 S(g_0^f,A_0^{\pr},\bar{A_0}^{\pr})$.

In \cite{ime} a detailed study of the gauged WZNW construction 
for  finite dimensional Lie algebras leading to Conformal NA Toda
models was presented. 
For an infinite dimensional  Kac-Moody
algebra  ${\widehat \lie }$
\br
[T_m^a ,T_n^b]  =  f^{abc} T^c_{m+n} + {\hat c}m \d_{m+n} \d^{ab} \nonu 
\er
\br
[{\hat d} , T^a_n] = nT^a_n ;\quad [{\hat c}, T^a_n] = [{\hat c},{\hat d} ] = 0
\label{km}
\er
the NA Toda models are associated to 
gradations of the type
$Q(h^{\pr}) = h^{\pr} d + 
\sum_{i\neq 1, 3 \cdots }^{r}\frac{2\lambda_{i}\cdot H^{(0)}}{\alpha_{i}^{2}}$, 
where $h^{\pr}$ is chosen
such that the gradation,
 $Q(h^{\pr})$, acting on
infinite dimensional Lie algebra $ \hat \lie$ ensures that the zero grade 
subgroup $\lie_0$ coincides with its
counterpart obtained with $Q(h^{\pr}=0)$ 
 acting on the Lie algebra $\lie $ of finite dimension apart from two  commuting 
 generators $\hat {c}$ and $\hat {d}$.  Since they commute with $\lie_0$,
  the kinetic part
 decouples such that the conformal and the affine singular NA-Toda  models 
 differ only by the
 potential term characterized by $ \eps_{\pm}$. 

The integration over the auxiliary gauge fields $A$ and $\bar A$ requires 
explicit
parametrization of $B$. 
\begin{eqnarray}
B=\exp ( {\sum_{l=1}^p \tilde \chi_{al}} E_{-\alpha_{al}}^{(0)})
 \exp (  \sum_{l=1}^{t} R^l \sum_{i=1}^{r}{{Y_i^l}} {{H_i^{(0)}}}+\Phi (H)+ \nu \hat {c} +
  \eta \hat {d})\exp (\sum_{l=1}^{p}\tilde \psi_{al} E_{\alpha_{al}}^{(0)})
 \label{63}
 \end{eqnarray}
where $ \Phi (H) 
=\sum_{j=1}^{r}\sum_{i=1}^{r-t}\varphi_{i}{{X}}_i^j {{H_j}^{(0)}}$,
 where $\sum_{j=1}^r{{Y_j^l}}  {{X^j_i}} =0, i=1, \cdots ,r-t, l=1, \cdots , t$.
After gauging away the nonlocal fields $R^l$, the factor group element becomes    
\be
g_0^f=\exp (\sum_{l=1}^{p}\chi_{al} E_{-\alpha_{al}}^{(0)})
 \exp (   \Phi (H)+ \nu \hat {c} + \eta \hat {d})\exp (\sum_{l=1}^{p}\psi_{al} E_{\alpha_{al}}^{(0)})
 \label{63a}
\ee
where $\chi_{al} = \tilde {\chi}_{al}e^{{1\o 2}\sum_{s=1}^t(Y^s\cdot \a_{al})R^s}, \quad 
\psi_{al} = \tilde {\psi}_{al}e^{{1\o 2}\sum_{s=1}^t(Y^s\cdot \a_{al})R^s}$.
We therefore get for the zero grade component
\br
F_0 &=&{-{k\o {2\pi}}}\int Tr\(  A_{0}\bar{\partial}g_0^f
(g_0^f)^{-1} + \bar{A}_{0}(g_0^f)^{-1}\partial g_0^f
+ A_{0} g_0^f\bar{A}_{0}(g_0^f)^{-1} + A_{0}\bar{A}_{0} \)d^2x \nonu \\
&=&{-{k\o {2\pi}}}\int ( \sum_{l^{\pr}, l^{\pr \pr}}^{t}a_{0l^{\pr }} 
\bar a_{0l^{\pr \pr}} 2(Y^{l^{\pr}}\cdot Y^{l^{\pr \pr }})
 \Delta_{l^{\pr} l^{\pr \pr}} \nonu \\
 &-&  \sum_{l, l^{\pr}}^{t}2({{\a_{al} \cdot Y^{l^{\pr}}}\o
{\a_{al}^2}})(\bar a_{0l^{\pr}}\psi_{al} \pa \chi_{al} + 
a_{0l^{\pr}} \chi_{al} \bar \pa \psi_{al} )e^{\Phi (\a_{al})})
d^2x 
\label{del}
\er
where $\Delta_{l^{\pr} l^{\pr \pr}} = 1 + 
\sum_{l}{{2 \o {\a_{al}^2}}}{{(Y^{l^{\pr}} \cdot \a_{al} )(Y^{l^{\pr \pr}} 
\cdot \a_{al} )}\o {2Y^{l^{\pr}}\cdot Y^{l^{\pr \pr}}}}\psi_{al} \chi_{al} e^{\Phi
(\a_{al} )}$ and 
$[\Phi (H), E_{\alpha_{al}}^{(0)}] = \Phi (\a_{al})E_{\alpha_{al}}^{(0)}$.   In matrix form $F_0$ reads,
\br
F_0 &=&{-{k\o {2\pi}}}\int \( \bar {\cal A}_0 M {\cal A}_0  - \bar {{\cal A}_0} N - \bar N {\cal A}_0 \)\nonu \\
&=&{-{k\o {2\pi}}}\int \( (\bar {{\cal A}_0} - \bar N M^{-1}) M (  {\cal A}_0 - M^{-1} N) - \bar N M^{-1} N \)
\label{delmatrgeneral}
\er
where
\br
{\cal A}_0 = \fourcol{a_{01}}{a_{03}}{ \vdots }{a_{0t}}, \quad \bar {\cal A}_0 =
\fourvec{\bar a_{01}}{\bar a_{03}}{\cdots }{\bar a_{0t}}, 
\quad 
N = \fourcol{\sum_{l=1}^{p} 2 {{\a_{al}\cdot Y^{1}}\o {\a_{al}^2}}\psi_{al} \pa \chi_{al} e^{\Phi(\a_{al})}}
{\sum_{l=1}^{p} 2 {{\a_{al}\cdot Y^{2}}\o {\a_{al}^2}}\psi_{al} \pa \chi_{al} e^{\Phi(\a_{al})}}{\vdots }{
\sum_{l=1}^{p} 2 {{\a_{al}\cdot Y^{t}}\o {\a_{al}^2}}\psi_{al} \pa \chi_{al} e^{\Phi(\a_{al})}} ,
\nonu
\er
\br
\bar N = \fourvec{\sum_{l=1}^{p} 2 {{\a_{al}\cdot Y^{1}}\o {\a_{al}^2}}\chi_{al} \bar \pa \psi_{al} 
e^{\Phi(\a_{al})}}{\sum_{l=1}^{p} 2 {{\a_{al}\cdot Y^{2}}\o {\a_{al}^2}}\chi_{al} \bar \pa \psi_{al} 
e^{\Phi(\a_{al})}}{\cdots }{\sum_{l=1}^{p} 2 {{\a_{al}\cdot Y^{t}}\o {\a_{al}^2}}\chi_{al} \bar \pa \psi_{al} 
e^{\Phi(\a_{al})}} ,\nonu
\er
and 
\br
M = \threemat{2{Y^1 \cdot Y^1}\Delta_{11}}{2{Y^1 \cdot Y^2}\Delta_{12}}{\cdots}{2{Y^2 \cdot Y^1}\Delta_{21}}{2{Y^2
 \cdot Y^2}\Delta_{22}}{\cdots }{\vdots }{\vdots }{\ddots}
 \label{mnn}
\er
The effective action is obtained by integrating over the auxiliary 
matrix fields ${\cal A}_0, \bar {\cal A}_0$, 
\be
Z_0 = \int D{\cal A}_{0}D\bar{{\cal A}}_{0}\exp (F_{0}) 
\ee  
The total effective action is therefore given as 
\br
S_{eff}&=&{-{k\o {2\pi}}}\int \( Tr \(  \pa \Phi (H) \bar \pa \Phi (H)\) +  2\sum_{l=1}^{p}\pa \chi_{al} \bar \pa \psi_{al}
e^{\Phi(\a_{al})} - \bar N M^{-1}N -V \)  
\label{action}
\er
where $V = Tr \( \eps_+ B \eps_- B^{-1}\) $.

\subsection{The Lund-Regge Model, $A_1^{(1)}$}

The simplest non abelian affine Toda model  consists in taking the 
$A_1^{(1)}$ Kac-Moody  algebra generated by $\{ E_{\pm \a}^{(n)}, H^{(n)} \} $, gradation
 $Q= d$ and $\eps_{\pm} =H^{(\pm 1)}$, such that $\lie_0 = SL(2)$. The 
 factor group $\lie_0 / \lie_0^0 = SL(2)/ U(1)$
 is then parametrized as 
\be
g_0^f=\exp (\chi E_{-\alpha}^{(0)})
  \exp (\psi E_{\alpha}^{(0)} )
 \label{g0f13}
\ee
We therefore find by direct calculation
\br
F_0 
={-{k\o {2\pi}}}\int \( a_{0} \bar a_{0} 2 \l_1   ^2\Delta
 -  \bar  a_{0} \psi \pa \chi  
 + a_0  \chi \bar \pa \psi  \)
d^2x 
\label{1dela32}
\er
where $\Delta = 1 +   \psi \chi   $.
After integration over $a_0$ and $\bar a_0$ we find the total effective 
action of the Lund-Regge model 
\br
{\cal L}_{eff} &=&   {{\pa \chi \bar \pa \psi }\o {\Delta}}   
 - (1+2\psi \chi )
\label{effac1}
\er
and equations of motion given by \cite{lundregge}
\br
\bar \pa \( {{\pa \chi }\o {\Delta }}\) + 
\chi {{\pa \chi \bar \pa \psi }\o {\Delta^2}} + 2 \chi &= &0, \nonu \\
 \pa \( {{\bar \pa \psi }\o {\Delta }}\) + 
\psi {{\pa \chi \bar \pa \psi }\o {\Delta^2}} + 2 \psi &= &0
\label{lreq}
\er

\subsection{The $p=1$, $A_2^{(1)}$ Model}

Consider the model defined by the  Kac-Moody algebra $A_2^{(1)}$, 
 gradation $Q = 2 \hat d + 2 {{\l_2 \cdot H} \o {\a_2^2}}$
 and constant grade $\pm 1$ generators $\eps_{\pm} =
E_{\pm \a_2}^{(0)} + E_{\mp \a_2}^{(\pm 1)}$. 
 The zero grade subspace is $SL(2) \otimes U(1)$,
generated by $\lie_0 = \{ E_{\pm \a_1}^{(0)}, h_1^{(0)}, h_2^{(0)} \}$. 
 It follows that 
$\lie_0^0 = \{ \l_1 \cdot H \} $, i.e. $Y^1 = \l_1$, 
$A_0 = a_0 \l_1 \cdot H, \;\;  \bar A_0 = \bar a_0 \l_1 \cdot H$.

The factor group element 
$g_0^f \in \lie_0 / \lie_0^0 = \( SL(2)\otimes U(1)\) / U(1)$ is then 
parametrized as 
\be
g_0^f=\exp (\chi E_{-\alpha_{1}}^{(0)})
 \exp \(   \varphi h_2^{(0)}\) \exp (\psi E_{\alpha_{1}}^{(0)} )
 \label{g0f12}
\ee
After integration over $a_0$ and $\bar a_0$ we find the total effective action
\br
{\cal L}_{eff} &=&   \pa \varphi \bar
\pa \varphi 
+ {{\pa \chi \bar \pa \psi }\o {\Delta}} e^{ - \varphi }  
 -\( e^{2\varphi } +e^{-2\varphi }  + \psi \chi e^{\varphi } \)
\label{effac2}
\er
where  $\Delta = 1 + {3 \o 4} \psi \chi e^{-\varphi }$, $\l_1^2 =  {2 \o 3}$. 
The equations of motion given by
\br
\bar \pa \( {{\pa \chi }\o {\Delta }}e^{-\varphi } \) + 
{3 \o 4}\chi {{\pa \chi \bar \pa \psi }\o {\Delta^2}}e^{-2\varphi } +  \chi e^{-\varphi } &= &0, \nonu \\
 \pa \( {{\bar \pa \psi }\o {\Delta }}e^{-\varphi } \) + 
{3 \o 4}\psi {{\pa \chi \bar \pa \psi }\o {\Delta^2}}e^{-2\varphi } +  \psi e^{-\varphi } &= &0 \nonu \\
\bar \pa \pa \varphi + {1\o 2} {{\pa \chi \bar \pa \psi }\o {\Delta^2}}e^{-\varphi } + 
e^{2\varphi } \( 1+ {1\o 2}\psi \chi
e^{-\varphi }\)- e^{-2\varphi }&= &0
\label{sl3}
\er


\subsection{The $p=t=2$, $A_3^{(1)}$ Model}

Consider the model defined by the  Kac-Moody algebra $A_3^{(1)}$, 
 gradation $Q = 2 \hat d + 2 {{\l_2 \cdot H} \o {\a_2^2}}$
 and constant grade $\pm 1$ generators $\eps_{\pm} =
E_{\pm \a_2}^{(0)} + E_{\mp \a_2}^{(\pm 1)}$. 
 The zero grade subspace is $SL(2)\otimes SL(2) \otimes U(1)$,
generated by $\lie_0 = \{ E_{\pm \a_1}, E_{\pm \a_3}, h_1, h_2, h_3\}$. 
 It follows that 
$\lie_0^0 = \{ \l_1 \cdot H, \l_3 \cdot H \} $, i.e. $Y^1 = \l_1, Y^2 = \l_3$.

The 2-singular structure is obtained by setting
\br
A_0 = a_{01}  \l_1 \cdot H + a_{03}  \l_3 \cdot H, \quad \bar A_0 = \bar a_{01}  \l_1 \cdot H + \bar a_{03} \l_3 \cdot H
\label{2bh}
\er
 The factor group element $g_0^f \in \lie_0 / \lie_0^0 = \( SL(2)\otimes SL(2) \otimes U(1) \) / U(1)^2$ 
 is then parametrized as 
\be
g_0^f=\exp (\chi_{1} E_{-\alpha_{1}}^{(0)}+ \chi_{3} E_{-\alpha_{3}}^{(0)})
 \exp (   \varphi h_2^{(0)})\exp (\psi_{1} E_{\alpha_{1}}^{(0)} + \psi_{3} E_{\alpha_{3}}^{(0)})
 \label{g0f2}
\ee
The total effective action is therefore given as 
\br
S_{eff}&=&{-{k\o {2\pi}}}\int  (  \pa \varphi \bar \pa \varphi + {{e^{-\varphi}}\o {2\Delta }} [ \Delta_3 \bar \pa \psi_1
\pa \chi_1 + {1\o 4} \chi_1 \psi_3 \bar \pa \psi_1 \pa \chi_3 e^{-\varphi} \nonu \\ 
&+& \Delta_1 \bar \pa \psi_3
\pa \chi_3  
+  {1\o 4} \chi_3 \psi_1 \bar \pa \psi_3 \pa \chi_1 e^{-\varphi} ] 
- V )
 \er
where $\Delta = {1\o 2} ( 1 + {3\o 4} (\psi_1 \chi_1 + \psi_3 \chi_3)e^{-\varphi} + 
 {1\o 2} \psi_1 \chi_1  \psi_3 \chi_3 e^{-2\varphi} )$, $\Delta_i = 1 + {3\o 4}\psi_i \chi_i e^{-\varphi}, i=1,3$ 
and 
$V= e^{-2\varphi} + e^{2\varphi}(1 + \psi_3 \chi_3 e^{-\varphi})(1 + \psi_1 \chi_1 e^{-\varphi})$.

The above examples describe non abelian Affine Toda Models 
characterized by broken conformal invariance and they all allow 
electrically charged soliton solutions.  For the models with
one singular structure, their spectra and soliton solutions where 
explicitly constructed in refs. \cite{dyonic} and
\cite{backlund}.  The generalized multi-charged model of example in this subsection
provides a generalization of the Lund-Regge model.  Its 
systematic  construction, its  soliton solutions and spectra  
are discussed in detail  in \cite{iraida}.

\sect{Zero Curvature and Equations of Motion }

The equations of motion for the action (\ref{spm}) are
\cite{lez-sav}
\be 
\bar \pa (B^{-1} \pa B) + [ {\eps_-}, B^{-1}  {\eps_+} B] =0, 
\quad \pa (\bar \pa B B^{-1} ) - [ {\eps_+}, B {\eps_-} B^{-1}] =0
\label{eqmotion}
\ee
 The subsidiary constraint $J_{Y^{l} \cdot H^{(0)}} =
  Tr(B^{-1} \pa B Y^{l}\cdot H^{(0)})$ and $
  \bar J_{Y^{l} \cdot H^{(0)}} = 
  Tr(\bar \pa B B^{-1}Y^{l} \cdot H^{(0)} )=        0$ can be
 consistently imposed  since $[Y^{l}\cdot H^{(0)},  {\eps_{\pm}}]=0, \;\; l=1, \cdots ,t \leq p$ as can be 
 obtained from
 (\ref{eqmotion}) by taking the trace with $Y^{l}\cdot H^{(0)}$.
We shall be considering $Y^{l} \cdot H = \l_{al} \cdot H, \;\; l=1, \cdots ,t$.
 In terms of parametrization (\ref{63a}), they yield the following system of equations for the nonlocal
 fields $R_k$
 \br
 \sum_{l=1}^{t} (\l_{ak} \cdot \l_{al} )\pa R_l  - 
  \tilde {\psi}_{al} \pa \tilde {\chi}_{al} e^{R_l - \Phi(\a_{al})} &=& 0 \nonu \\
 \sum_{l=1}^{t} (\l_{ak} \cdot \l_{al} )\bar \pa R_l  - 
  \tilde {\chi}_{al} \bar \pa \tilde {\psi}_{al} e^{R_l - \Phi(\a_{al})} &=& 0
 \label{bh1}
 \er
 which can be solved in terms of new variables $\psi_{al} = \tilde {\psi}_{al}e^{{1\o 2}R_l}, \;\; 
 \chi_{al} = \tilde {\chi}_{al}e^{{1\o 2}R_l}$.  Since $\l_j = K^{-1}_{jl} \a_{l}$, 
 we rewrite (\ref{bh1}) in matrix form as
 \br
 \sum_{l=1}^{t} \( K^{-1}_{aj, al} + {1\o 2} 
 {\psi}_{al}{\chi}_{al}e^{-\Phi(\a_{al})} \d_{aj, al}\) \pa R_{l} &=& 
 \psi_{aj}\pa {\chi}_{aj}e^{-\Phi(\a_{aj})}, \nonu \\
 \sum_{l=1}^{t} \( K^{-1}_{aj,al} + {1\o 2} 
 {\psi}_{al}{\chi}_{al}e^{-\Phi(\a_{al})} \d_{aj,al}\) \bar \pa R_{l} &=& 
 \chi_{aj}\bar \pa {\psi}_{aj}e^{-\Phi(\a_{aj})}, 
 \;\; j=1, \cdots ,t
 \label{bh2}
 \er
 In matrix form, eqns. (\ref{bh2}) can be rewritten as
 \br
 M_{ij} \pa R_{aj} = N_{i}, \quad  M_{ij} \bar \pa R_{aj} = {{\bar N}^{tr}}_{i} 
 \label{matrr}
 \er
 where $M, N$ and $\bar N$ are given in (\ref{mnn}). 
 The  solution is of the form
 \br
 \pa R_{ai} = \( M^{-1}\)_{ij} \( N \) _{j}, \quad 
 \bar  \pa R_{ai} = \( M^{-1}\)_{ij} \( {\bar {N}^{tr}} \) _{j}\nonu
 \er

Alternatively, (\ref{eqmotion}) admits a
zero curvature representation
\br 
\pa \bar A - \bar \pa A - [A, \bar A] =0
\label{zcca}
\er
 where
\be
A= -B  {\eps_-}  B^{-1} ,\quad  
\bar A=  {\eps_+}   + \bar \pa B B^{-1} 
\label{zcc}
\ee
Whenever the constraints (\ref{bh2}) are incorporated into $A$ and $\bar A$ in
(\ref{zcc}), equations (\ref{zcca}) yield the equations of motion  of the NA singular Toda models, which coincide with those
derived from the general action (\ref{action}).

\section{Multi-time Evolution Equations}

In this section we review the construction of the cKP hierarchy 
in terms of the decomposition of zero grade algebra 
$\lie_0 = Im  \( ad \eps_+ \) \oplus  Ker \( ad \eps_+ \)$ \cite{jmp}.
 The time evolution equations   are given in terms 
of the zero curvature representation, 
\br
\pa _{t_N} A_{x} - \pa_{x} A_{t_N} - [ A_{t_N}, A_{x} ] =0, \quad N > 0
\label{postime}
\er
where $A_{x}= A_0 + \eps_+$, $A_0 \in \lie_0$, $A_0 \in Im  \( ad \eps_+ \)$ and 
\br
A_{t_N} = D_{N}^{(N)} + D_{N}^{(N-1)} + \cdots D_{N}^{(0)}, \quad  D_{N}^{(k)} \in \lie_{k}
\label{bN}
\er
where the upper index denotes the grading, i.e. $D_N^{(j)} \in \lie_j$.
Equation (\ref{postime}) can be decomposed according to the grading into the system of algebraic equations 
\br
[D_N^{(N)}, \eps_+ ] &=& 0 \nonu \\
-\pa_{x} D_N^{(N)}  - [D_N^{(N)}, A_0 ] - [D_N^{(N-1)}, \eps_+ ] &=& 0 \nonu \\
-\pa_{x} D_N^{(N-1)}  - [D_N^{(N-1)}, A_0 ] - [D_N^{(N-2)}, \eps_+ ] &=& 0 \nonu \\
\vdots &=& \vdots \nonu \\
-\pa_{x} D_N^{(1)}  - [D_N^{(1)}, A_0 ] - [D_N^{(0)}, \eps_+ ] &=& 0 \nonu \\
\pa_{t_N}A_0 -\pa_{x} D_N^{(0)}  - [D_N^{(0)}, A_0 ] &=& 0
\label{system}
\er
which can be solved by  taking $ D_N^{(N)} \in Ker \( ad \eps_+ \)$. 
 Substituting the result in the second eqn. (\ref{system}) we  determine 
$ D_N^{(N)}$ and the component of $ D_N^{(N-1)}$ lying in the $Im \( ad \eps_+ \)$. 
Substituting in the third eqn. (\ref{system}) we  determine 
the component of $ D_N^{(N-1)}$ lying in the $ Ker \( ad \eps_+ \)$
and the component of $ D_N^{(N-2)}$ lying in the $Im \( ad \eps_+ \)$.  We solve for 
$ D_N$ by repeating the argument 
$N$ times until we reach the last eqn. (\ref{system}) from where we determine the 
component of $ D_N^{(0)}$ lying in the $ Ker \( ad \eps_+ \)$ together with the 
$t_{N}$-evolution equations for the fields $A_0 \in Im \( ad \eps_+ \)$.

As examples of positive grade time evolution let us consider the first nontrivial case $t_{N}=t_2 $.

\subsection{AKNS Hierarchy}
The $A_1^{(1)}$ AKNS hierarchy is defined by  
\br
A_0 = qE_\a^{(0)} + r E_{-\a}^{(0)}, \quad  \eps_+ =  H^{(1)} = \mu \twomat{1}{0}{0}{-1}
\er
where $\mu $ denote the spectral parameter.
The equations of motion for $t_{N}=t_2$ in (\ref{system}) are given 
by the non linear Schroedinger equation,
\br
\pa_{t_2} q + \pa^2_x q     -2r q^2  &=&0 \nonu \\
\pa_{t_2} r - \pa^2_x r  + 2r^2 q  &=&0 
\label{eqnt2akns}
\er

\subsection{Yajima-Oikawa Hierarchy}
The $A_2^{(1)}$ hierarchy is defined by the following algebraic structure
\br
A_0 = qE_{\a_1}^{(0)} + r E_{-\a_1}^{(0)}+ Uh^{(0)}_2 , \quad  \eps_+ =  E_{\a_2}^{(0)} + E_{-\a_2}^{(1)}
\er
and the equations of motion obtained from (\ref{system}) with $t_{N}=t_2$ are   the Yajima-Oikawa equations, 
\br
\pa_{t_2} q - \pa^2_x q + q\pa_x   U   + q U^2 + r q^2   &=&0 \nonu \\
\pa_{t_2} r + \pa^2_x r + r \pa_x  U   - r U^2 - r^2 q  &=&0 \nonu \\
\pa_{t_2} U + \pa_x \( r q \)  &=&0
\label{eqnt2yo}
\er

\subsection{The  $p=t=2$  $A_3^{(1)}$ Hierarchy}

Let  $\lie = \hat {SL(4)}$ hierarchy defined by 
\br
A_0 = q_1 E_{\a_1}^{(0)}+ q_3 E_{\a_3}^{(0)} + 
r_1 E_{-\a_1}^{(0)}+ r_3 E_{-\a_3}^{(0)} + Uh_2^{(0)}, \quad \eps_+ = E_{\a_2}^{(0)}+ E_{-\a_2}^{(1)}
\er
  The $t_2$ time 
evolution is then written as 
\br
\pa_{t_2} q_1 - \pa^2_x q_1 + q_1\pa_x   U   + q_1 U^2 + r_1 q_1^2 + r_3 q_1 q_3 &=&0 \nonu \\
\pa_{t_2} r_1 + \pa^2_x r_1 + r_1 \pa_x  U   - r_1 U^2 - r_1^2 q_1 - r_1 r_3 q_3 &=&0 \nonu \\
\pa_{t_2} q_3 + \pa^2_x q_3 - q_3 \pa_x  U   - q_3 U^2 - r_3 q_3^2 - r_1 q_1 q_3 &=&0 \nonu \\
\pa_{t_2} r_3 - \pa^2_x r_3 - r_3\pa_x  U   + r_3 U^2 + r_3^2 q_3 + r_3 r_1 q_1 &=&0 \nonu \\
\pa_{t_2} U + \pa_x \( r_1 q_1 \) -  \pa_x \( r_3 q_3 \) &=&0
\label{eqnt2}
\er

\sect{Negative grade  Evolution Equations}

The relativistic invariant models  constructed in section 2 
 share  the same algebraic structure of the 
Constrained KP (cKP) models discussed  in ref. \cite{jmp}.  
As explained in the examples, the Lund-Regge and its $SL(3)$ generalization 
 belong to the non linear Schroedinger
and the Yajima-Oikawa hierarchies  respectively. 
The $SL(4)$ example  of subsections  $(2.3)$ and $(4.3)$ also correspond to the same hierarchy.
Their field transformation  is given by 
\br
A_0 = \bar \pa B B^{-1}
\label{5.1}
\er
and its soliton solution are related.
In fact the Leznov-Saveliev equations (\ref{eqmotion}) correspond to 
zero curvature equation for $t_{-1} = z$. It was shown
in \cite{jpa7} that the general negative ``time'' hierarchy is given by 
\br
\pa _{t_{-N}} A_{x} - \pa_{x} A_{t_{-N}} - [ A_{t_{-N}}, A_{x} ] =0, \quad N > 0
\label{negtime}
\er
where 
\br
A_{t_{-N}} = D_{N}^{(-N)} + D_{N}^{(-N+1)} + \cdots D_{N}^{(-1)}, \quad  D_{N}^{(-k)} \in \lie_{-k}
\label{-bN}
\er
For $N=1$, a general  solution is given in closed  form by 
\br
D_{1}^{(-1)} =- B\eps_- B^{-1}
\er

\sect{One Soliton Solutions}
In this section we discuss the relation between the one soliton solutions within the same hierarchy.
Let us first consider the $Sl(2)$ case of the Lund-Regge equations of motion (\ref{lreq})
 with solution given by 
 \br
 \psi &=& {{b e^{{{z}\o {\g_2}}+ \bar z \g_2}}\o {1+\Gamma e^{z ({1\o {\g_2}} - {1\o {\g_1}})+ \bar z (\g_2- \g_1)}   }}, 
 \quad  
 \chi = {{a e^{-{{z}\o {\g_1}}- \bar z \g_1}}\o {1+\Gamma e^{{{z}({1\o {\g_2}} - 
 {1\o {\g_1}})}+ \bar z (\g_2- \g_1)} }}, \nonu \\
 e^{R} &=& {{1+{{\g_1} \o {\g_2}}\Gamma e^{{{z}({1\o {\g_2}} - {1\o {\g_1}})}+ \bar z (\g_2- \g_1)}}\o 
 {1+\Gamma e^{{{z}({1\o {\g_2}} - {1\o {\g_1}})}+ \bar z (\g_2- \g_1)}}} \nonu \\
 \label{sol1}
\er 
where $
\Gamma = {{ab \g_1 \g_2}\o {(\g_1 -\g_2 )}}, \quad z = t_{-1}, \;\; \bar z = x$

The one soliton solution of the nonlinear Schroedinger equation (\ref{eqnt2akns}) is given by
\br
r = - {{a \g_1 e^{-\g_1^2 t_2 - \g_1 x} }\o { 1+ {{ab \g_1 \g_2 }\o {(\g_1 -\g_2 )^2}}e^{(\g_2^2 -\g_1^2)t_2
+ (\g_2 -\g_1)x }}}, \quad 
q =  {{b \g_2 e^{\g_2^2 t_2 + \g_2 x} }\o { 1+ {{ab \g_1 \g_2 }\o {(\g_1 -\g_2 )^2}}e^{(\g_2^2 -\g_1^2)t_2
+ (\g_2 -\g_1)x }}} 
\label{sol2}
\er
The solutions agree when we parametrize $q$ and $r$ in terms of $\psi$ and $\chi$ defined as in  (\ref{5.1}), i.e.
\br
q= {{\bar \pa \psi }\o {\Delta}}e^{R}, \quad 
r= {{\bar \pa \chi }}e^{-R}
\er
and ${z \o \g_i} \rightarrow \g_i^2 t_2, i=1,2, \; \bar z \rightarrow x $.

For the $Sl(3)$ case  the solution of the relativistic model was given in ref. \cite{dyonic} as 
\br
e^{{2\o 3} R} &=& {{1+ {{ab \rho_1 \rho_2 }\o {(1 - \g_{1,2})(1- \g_{1,2}^2)}}\o 
{1+ {{ab \rho_1 \rho_2 \g_{1,2}^2}\o {(1 - \g_{1,2})(1- \g_{1,2}^2)}}}}}, \quad 
e^{\varphi } ={{1+ {{ab \rho_1 \rho_2 \g_{1,2}}\o {(1 - \g_{1,2})(1- \g_{1,2}^2)}}}\o {
{{(1+ {{ab \rho_1 \rho_2 \g_{1,2}^2}\o {(1 - \g_{1,2})(1- \g_{1,2}^2)}})} 
\( {A + ab \rho_1 \rho_2 }\o {A + ab \rho_1 \rho_2\g_{1,2}^2 }\)^{{1/2}}}}}\nonu \\
\chi &=& {{Aa\rho_1 }\o {(A + ab \rho_1 \rho_2 )}^{{1\o 4}} (A + ab \rho_1 \rho_2
\g_{1,2}^2)^{{3\o 4}}} \nonu \\
\psi &=& {{Ab\rho_2 }\o {(A + ab \rho_1 \rho_2 )}^{{1\o 4}} (A + ab \rho_1 \rho_2
\g_{1,2}^2)^{{3\o 4}}}
\label{relyo}
\er
where 
\br
\rho_1 = e^{-{{ z }\o {\g_1}} +  \bar z \g_1}, \quad  \rho_2 = e^{{{ z }\o {\g_2}} -  \bar z \g_2}, \quad
\g_{1,2} = \g_1 / \g_2, \quad  A = (1 -\g_{1,2} )(1- \g_{1,2}^2)
\er
It therefore follows from  (\ref{5.1}) that 
\br
q= {{ \bar \pa \psi }\o {\Delta }}e^{{1\o 2}R - \varphi }, \quad 
r=  \( \bar \pa \chi - \chi \bar \pa \varphi - {1\o 4}
{{\chi^2 \bar \pa \psi e^{-\varphi}}\o {\Delta }} \) e^{{-1\o 2}R  }, \quad 
U = \bar \pa \varphi + {{\chi \bar \pa \psi e^{-\varphi}}\o {2\Delta }}
\label{qru}
\er
where $\Delta = 1 + {3\o 4} \psi \chi e^{-\varphi}$.

Substituting (\ref{relyo}) in (\ref{qru})  we find that the relativistic solution agrees 
with the solution of the Yajima-Oikawa model given in \cite{aratyn}
\br
q&=&  -{{b\g_2e^{\g_2^2t_2 -x\g_2}}\o {1+ {{ab\g_2\g_1^2 e^{x(\g_1 -\g_2) - 
t_2( \g_1^2- \g_2^2)}}\o {(\g_1+\g_2)(\g_1-\g_2)^2}}}}, \quad
 r =  {{a\g_1e^{-\g_1^2t_2 +x\g_1}}\o {1+ {{ab\g_2^2\g_1 e^{x(\g_1 -\g_2) - t_2( \g_1^2- \g_2^2)}}\o
 {(\g_1+\g_2)(\g_1-\g_2)^2}}}},\nonu \\
 U&=& -\pa_x ln \( {{1+ {{ab\g_2\g_1^2 e^{x(\g_1 -\g_2) - 
t_2( \g_1^2- \g_2^2)}}\o {(\g_1+\g_2)(\g_1-\g_2)^2}}} \o {1+ {{ab\g_2^2\g_1 e^{x(\g_1 -\g_2) - t_2( \g_1^2- \g_2^2)}}\o
 {(\g_1+\g_2)(\g_1-\g_2)^2}}}}
 \)
\er
 after the substitution
${z \o \g_i} \rightarrow \g_i^2 t_2, i=1,2, \; \bar z \rightarrow x$.

\sect{Concluding Remarks}

The relation between certain relativistic non abelian Toda  models constructed 
in this notes as the negative flows of the constrained
KP (generalized non linear Schroedinger, Yajima-Oikawa, etc ) 
non relativistic models is shown to generalize the well known relation
between the sine-Gordon and mKdV  or the Lund-Regge and non linear Schroedinger
models.  The common  algebraic structure is the key to propose negative grade
 hierarchies containing the relativistic equations 
as the first negative flows.  An interesting question that  naturally arises 
concerns the  relationship between the conserved charges. In order to illustrate 
that 
 let us consider for instance the Yajima-Oikawa model where 
the use of eqns. of motion (\ref{eqnt2yo})  yields
\br
\pa_{t_2} \( U^2 + rq \) = \pa_x \( r\pa_x q - q \pa_x r -2 rqU \)
\label{hamyo}
\er
and hence the first hamiltonian density
\br
{\cal {H} }_{non \; rel}= rq +U^2
\er
Substituting (\ref{qru}) in (\ref{hamyo}) we find that the hamiltonian density 
of the $SL(3), p=1$ model described in subsection {\bf II.B}, 
\br
{\cal {H}}_{ rel} =  {{\bar \pa \psi \bar \pa \chi }\o 
{\Delta }}e^{-\varphi } + (\bar \pa \varphi )^2, \quad 
\Delta = 1+ {3\o 4}\psi \chi e^{-\varphi}
\er
is also conserved due to eqns. of motion (\ref{sl3}), i.e.
\br
\pa_{t_{-1}} {\cal {H}}_{ rel}=  \bar \pa \( \psi \chi e^{-\varphi} \)
\er
with the identification $t_{-1}=z, x = \bar z$.
Similar relations can be found for higher grade equations within 
the same hierarchy.

\vskip 10pt \noindent
{\bf Acknowledgements} \\
We  thank H. Aratyn for discussions on the negative flow formulation and Fapesp, Unesp and CNPq for support.

\end{document}